\documentclass[aps,prl,twocolumn,superscriptaddress]{revtex4}

\usepackage{amsmath}
\usepackage{amssymb}
\usepackage{graphicx}

\newcommand{\vphi}{\varphi}
\newcommand{\eps}{\varepsilon}
\newcommand{\nn}{\nonumber}

\begin{document}

\title{Phase separation transition in liquids and polymers induced by electric
field gradients}

\author{Gilad Marcus}
\affiliation{Max-Planck-Institut f\"{u}r Quantenoptik, Hans-Kopfermann-Str. 1,
D-85748
Garching, Germany.}

\author{Yoav Tsori}
\affiliation{Department of Chemical Engineering, Ben-Gurion University of the
Negev,
84105 Beer-Sheva, Israel.}

\begin{abstract}

Spatially uniform electric fields have been used to induce instabilities
in liquids and polymers, and to orient and deform ordered phases of
block-copolymers. Here we discuss the demixing
phase transition occurring in liquid mixtures when they are subject to spatially
{\it nonuniform} fields.
Above the critical value of potential, a phase-separation transition occurs,
and two coexisting phases appear separated by a sharp interface.
Analytical and numerical composition profiles are given, and the 
interface location
as a function of charge or voltage is found. The possible influence
of demixing on the stability of suspensions and on inter-colloid interaction
is discussed.

\end{abstract}

\maketitle

\section{Introduction}

Electric fields influence the structure and thermodynamic behavior
of charged as well as neutral matter. Their effect is strong, they
can be switched on or off, and they are easily scalable to the
sub-micron regime \cite{LL1}. There are two main distinctions with
respect to the field: spatially uniform vs. nonuniform fields. There
are also two broad classes of material properties: pure dielectric
vs. conducting media. All four combinations are relevant to phase
transitions in liquid and polymer mixtures and to liquid-vapor
coexistence in pure liquids.

{\bf Perfect dielectrics.} The electrostatic energy of dielectric materials is
given by the expression
\begin{equation}\label{f_es}
F_{\rm es}=-\frac12\int\eps{\bf E}^2{\rm d}^3r
\end{equation}
where $\eps$ is the dielectric constant and ${\bf E}$ is the
electric field. The negative sign before the integral is applicable to
situations where
the electric potential $\psi$ (${\bf E}=-\nabla\psi$) is given on the
bounding surfaces; in cases where the charge is prescribed,
${\bf E}$ is given as a function of the displacement field ${\bf D}$, and the Legendre transform
reverses the sign \cite{LL1}.

The phase-transition
described below occurs in systems described by bistable free-energy
functionals giving rise to a phase-diagram in the
composition-temperature plane divided into two regions: homogeneous
mixture and a phase-separated state. For concreteness,
we consider the symmetric mixture free-energy density $f_m$ given by
\begin{eqnarray}\label{fm}
&&\frac{v_0}{k_BT}f_{m}=\\
&&\left[\phi\log(\phi)+
(1-\phi)\log(1-\phi)\right]+2k_BT_c\phi(1-\phi)\nn
\end{eqnarray}
This free-energy is given in terms of the dimensionless composition
$\phi$ ($0\leq\phi\leq 1$). In a binary mixture of two liquids 1 and
2, with dielectric constants $\eps_1$ and $\eps_2$, $\phi$ is the
relative composition of (say) liquid 2. In A/B polymer blends, it is
the relative volume fraction of polymer A, and similarly for an A/B
diblock-copolymer melt. $v_0$ is a molecular volume, $k_B$ is the
Boltzmann constant and $T_c$ is the critical temperature. 
In symmetric mixtures, the
transition (binodal) temperature $T_t$ at a given composition is
given by $df_m(T_t,\phi)/d\phi=0$, that is, 
$T_t/T_c=2(2\phi-1)/\log(\phi/(1-\phi))$ \cite{doi}.
 In the absence of electric field, the
mixture is homogeneous if $T>T_t$, and unstable otherwise.
The field-induced phase-transition discussed below
does not depend on the exact form of $f_m$; it occurs
also in a Landau series expansion of Eq. (\ref{fm}) around
the critical composition
$\phi_c$, and in other forms having a ``double-well'' shape.

The dielectric constant $\eps$ depends on $\phi$ via a constitutive
relation. A variation of $\phi$ from its critical value, $\phi_c$,
induces a variation of $\eps$ from the critical permittivity
$\eps_c$. When the composition deviation $\vphi\equiv\phi-\phi_c$ 
is small
enough, $|\vphi|\ll 1$, the constitutive relation $\eps(\phi)$
can be written as a Taylor series expansion to quadratic order:
\begin{eqnarray}\label{const_relation}
\eps(\phi)=\eps_c+\Delta\eps\vphi+\frac12\eps''\vphi^2
\end{eqnarray}
The ``dielectric contrast'' $\Delta\eps$ is
simply equal to $\eps_2-\eps_1$, if $\eps''$ vanishes.

The electric field depends on the imposed external potentials or charges and
on the local dielectric constant.
Let us denote by ${\bf E}_0$ the
electric field corresponding to the system with uniform composition $\phi_c$
everywhere.
Composition changes in $\phi$ induce changes in $\eps$, and
since $\eps$ and ${\bf E}$ are coupled via Laplace's equation
$\nabla(\eps{\bf E})=0$,
one has variations in electric field.

We may thus write to quadratic order in $\vphi$
\begin{eqnarray}\label{E_expansion}
{\bf E}={\bf E}_0+{\bf E}_1\vphi+\frac12{\bf E}_2\vphi^2~~.
\end{eqnarray}
Note that ${\bf E}_0$ is constant in space only inside a parallel-plate
capacitor, or if the sources of the field are very far from the system under
investigation. Clearly, even if ${\bf E}_0$ is uniform,
composition variations lead to field nonuniformities.

One can expand the electrostatic energy density in
Eq. (\ref{f_es}) in powers of $\vphi$:
\begin{eqnarray}\label{f_es_expansion}
f_{\rm es}&=&const.-\left(\eps_c{\bf E}_0\cdot{\bf
E}_1+\frac12\Delta\eps{\bf E}_0^2\right)\vphi\\
&-&\frac12\left(\frac12\eps''{\bf E}_0^2+
\eps_c{\bf E}_1^2+2\Delta\eps{\bf E}_0\cdot{\bf
E}_1+\eps_c{\bf E}_2\cdot{\bf
E}_0\right)\vphi^2\nn\\
&+&O(\vphi^3)~~.\nn
\end{eqnarray}
The unimportant constant corresponds to the electrostatic energy of the system
with uniform composition, and it serves as a reference energy.
If the field ${\bf E}_0$ is uniform in space,
the two terms in linear order of
$\vphi$ simply add a constant to the chemical potential, and therefore are
inconsequential for the
thermodynamic state of the system \cite{footnote}.

Landau and Lifshitz showed that the existence of a $\eps''\vphi^2$ term
in Eq. (\ref{f_es_expansion})
is responsible to a shift
of the critical temperature $T_c$ \cite{LL1,tsori_rmp2009}. They
found that $T_c$ is increased by $\Delta T_c$ given by
\begin{eqnarray}
\Delta T_c=\frac{v_0\eps''E_0^2}{2k_B}~~~,
\label{DT_landau_mechanism}
\end{eqnarray}
$T_c$ and the whole binodal
curve close to the critical point are increased if $\eps''>0$
(field-induced demixing) or decreased if $\eps''<0$
(field-induced mixing). A similar expression exists for a pure liquid
in coexistence with its vapor. 

The experiments, starting with P. Debye and Kleboth
\cite{debye_jcp1965}, are in contradiction with this prediction.
Debye and Kleboth investigated the critical temperature of a
Isooctane-Nitrobenzene mixture (relative permittivities $2.0$ and
$34.2$, respectively) They observed reduction of $T_c$ by $15$ mK in
a field of $4.5$ V/$\mu$m. Their measurements were later verified by
Orzechowski \cite{orzech_chemphys1999}. Beaglehole worked with a
Cyclohexane-Aniline mixture (relative permittivities $2$ and $7.8$,
respectively), and he measured reduction of $T_c$ by as much as $80$
mK in a $0.3$ V/$\mu$m dc field \cite{beaglehole_jcp1981}. Early
worked on the same mixture but in $1$ V/$\mu$m ac field, and found
no change in $T_c$. He attributed the results of Beaglehole to
spurious heating \cite{early_jcp1992}. Wirtz and Fuller performed
similar experiments on n-hexane-Nitroethane mixture (relative
permittivities $2$ and $19.7$, respectively), and found a reduction
of $T_c$ by $20$ mK in a $5$ V/$\mu$m field
\cite{wirtz_fuller_prl1993}. In all cases, $\eps''$ was positive but
still mixing was observed. In addition, the observed change in $T_c$ is quite
small, typically in the $10$-$20$ mK range. The only exception is the work
of Gordon and Reich \cite{reich_jpspp1979}. They worked on polymer
mixtures of poly(vinyl methyl ether) (PVME)-polystyrene (PS) system
(relative permittivities $2.15$ and $2.6$, respectively), and
observed changes significantly larger than $1$ K. Their strong
effect can be attributed to the large molecular weight of the
polymer ($14,000$ -- $30,000$ gr/mol) and to their reduced entropy compared
to that of simple liquids.

One is inclined to explain the experimental findings by the
second and third terms on the second line of Eq. (\ref{f_es_expansion})
(proportional to $\vphi^2$).
The third term is
twice as large as the second one and opposite in sign, and the two sum to give a
free energy
contribution proportional to the dielectric contrast squared $+(\Delta\eps)^2$. This is a
free energy penalty for dielectric interfaces perpendicular to the external
field.
Indeed, these additional terms are responsible
to the normal field instability in liquids \cite{tsori_rmp2009},
and to orientation of ordered phases
(e.g. block-copolymers) in external fields
\cite{ah_mm1991,ah_mm1993,ah_mm1994,russell_sci1996,muthu_jcp2001,onuki_mm1995,TA_mm2002}.
In liquid mixtures they favor mixing (lowering of $T_c$).

\section{Mixtures of nonpolar liquids in fields gradients}

Field gradients are general, and occur in all electrodes unless
special care is taken to eliminate them (super-flat and parallel
conducting surfaces). When mixtures of pure dielectric liquids are
subjected to a {\it spatially nonuniform field},
the situation is very different. The direct coupling between field
variations and composition fluctuations then leads to a
dielectrophoretic force, depending on $\Delta\eps$ in Eq.
(\ref{const_relation}), which tends to ``suck'' the component with
large $\eps$ to regions with high electric field, as in
the case of the well-known rise of a dielectric liquid in a
capacitor \cite{LL1}.
\begin{figure}[h!]
\begin{center}
\includegraphics[scale=0.5,bb=55 580 530 710,clip]{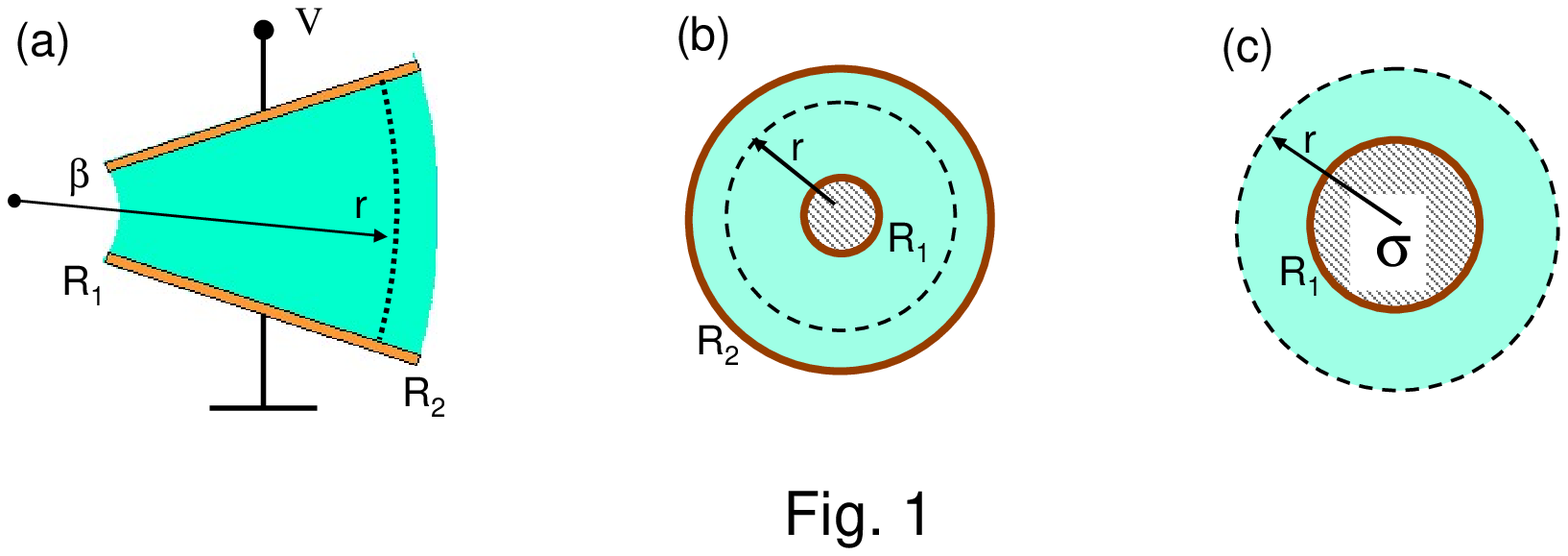}
\end{center}
\caption{Three model systems where field gradients lead to
demixing.
(a) A wedge comprised of two flat electrodes with an opening angle $\beta$ and
potential
difference $V$. $R_1$ and $R_2$ are the minimal and maximal values of the
distance $r$
from the imaginary meeting point.
(b) A charged wire with radius $R_1$, or two concentric cylinders 
with radii $R_1$ and $R_2$.
(c) A single charged
colloid of radius $R_1$ and surface charge $\sigma$.
}
\label{fig_3geometries}
\end{figure}

~\\
{\bf Statics}\\
Three ``canonical'' geometries with electric field gradients are presented in
Fig. \ref{fig_3geometries}.
The first is the
``wedge'' capacitor, made up from two flat and nonparallel surfaces with potential
difference $V$, and opening angle $\beta$. The electric field is then azimuthal,
${\bf E}(r)=V/(\beta r)\hat{\bf \theta}$, where $r$ is the distance from the imaginary
meeting point of the surface. $r$ is bounded by the smallest and largest
radii $R_1$ and $R_2$, respectively. The second model system consists of a
charged wire of radius $R_1$, or two concentric metallic cylinders with radii
$R_1$ and $R_2>R_1$.
In this
case the azimuthally-symmetric field is
${\bf E}(r)=\sigma R_1/(r\eps(r))\hat{\bf r}$, where $\sigma$ is the
charge per unit area on the inner cylinder. Lastly, for a charged spherical
colloid of radius $R_1$ and surface charge $\sigma$, one readily finds the
spherically-symmetric field to be
${\bf E}(r)=\sigma R_1/(r^2\eps(r))\hat{\bf r}$.
\begin{figure}[h!]
\begin{center}
\includegraphics[scale=0.55,bb=140 265 455 775,clip]{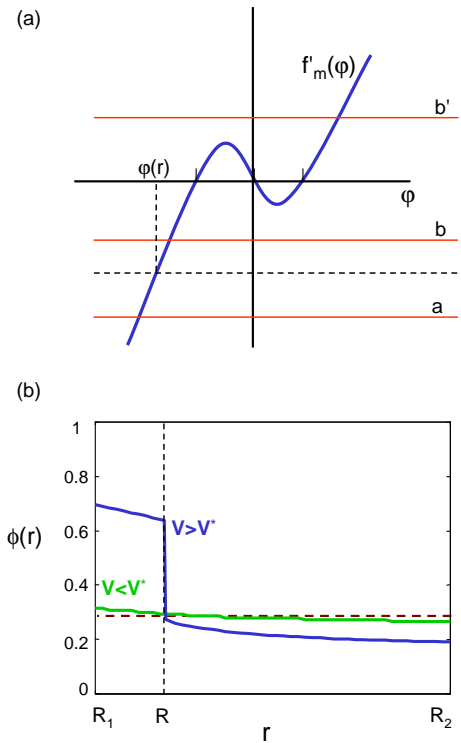}
\end{center}
\caption{(a) Graphical solution to Eq. (\ref{wedge_gov_eqn}). Solid curve is $f_m'(\vphi)$.
Its roots are the transition (binodal) compositions.
The intersection between $f_m'(\vphi)$ and the horizontal dashed line gives the solution
$\vphi(r)$ to Eq. (\ref{wedge_gov_eqn}). For voltages $V$ below the critical value $V^*$,
the dashed line is bounded by lines a and b, corresponding to the the maximal
and minimal values
of the right-hand side of Eq. (\ref{wedge_gov_eqn}), giving rise to a continuous profile
$\vphi(r)$. At $V>V^*$, line b is displaced to b', and the intersection is at $\vphi<0$
for large $r$'s and at $\vphi>0$ at small $r$'s. (b) Qualitative composition profiles $\phi(r)$.
Horizontal dashed line is the average composition $\phi_0$ in the absence of field.
$\phi(r)$ varies smoothly when $V<V^*$, and has a sharp jump at $r=R$ when $V>V^*$.
}
\label{fig_comp_profiles}
\end{figure}

In all three cases a general scenario occurs: when $T$ is
above $T_c$, the composition profile $\phi({\bf r})$ is smooth, and its
gradients increase as the charge on the objects increases. However, below $T_c$
the behavior is different -- $\phi({\bf r})$ is smooth as long as the charge
(voltage) is small, and becomes {\it discontinuous} when the charge (voltage)
attains a critical value. At this charge, a sharp interface appears between
the coexisting domains \cite{TTL_nature2004,KYL}.
As the charge further increases, the
interface location and the compositions of the coexisting
domains change.

To see this, consider the wedge capacitor, for which the electrostatic energy
density is
\begin{eqnarray}
f_{\rm es}=-\frac12\left(\eps_c+\Delta\eps\vphi\right)\left(\frac{V}
{\beta r}\right)^2
\end{eqnarray}
Note that we have used a linear constitutive relation. In uniform electric fields,
such a linear relation would mean that the electrostatic energy is simply a constant
independent of the composition profile. In addition, in the wedge geometry
the electrostatic energy does not have a term proportional to $(\Delta\eps)^2$
because the electric field is parallel to the dielectric interfaces
(both are in the $\hat{\bf \theta}$ direction).

The equation that governs the composition profile $\vphi(r)$,
derived from the Euler-Lagrange equation $\delta(f_m + f_{\rm es})/\delta\varphi =0$,
is the following:
\begin{eqnarray}\label{wedge_gov_eqn}
f_m'(\vphi)=\frac12\Delta\eps\left(\frac{V}{\beta r}\right)^2+\mu~~.
\end{eqnarray}
Here $\mu$ is the chemical potential of the large reservoir at infinity.
Note that Eq.
(\ref{wedge_gov_eqn}) gives an analytical expression for $r$ as a
function of $\vphi$.

The right hand side of the equation is independent of $\vphi$, and is indicated
by the horizontal lines in Fig. \ref{fig_comp_profiles} (a). Suppose the mixture
composition in the absence of field corresponds to a point above and to the left
of the binodal curve (homogeneous mixture).
A graphical solution of the governing equation is obtained by the
intersection of the horizontal line, whose location depends on the field, and
therefore on $r$, with the curve $f_m'(\vphi)$.
If $T$ is above $T_c$, $f_m$ is convex, and therefore the intersection of the two
curves changes smoothly as $r$ decreases ($E$ increases). The resulting
composition profile is shown in Fig. \ref{fig_comp_profiles} (b).

However, the situation is different below $T_c$: here $f_m'(\vphi)$
behaves like $-\vphi+\vphi^3$. When the applied voltage is small
enough such that the maximum value of the right-hand side of Eq.
(\ref{wedge_gov_eqn}) occurs at line b of Fig.
\ref{fig_comp_profiles} (a), as one goes from large to small values
of $r$ (increasing $E$), the composition increases, but $\vphi(r)$
is always continuous. There is a critical value of the voltage,
$V^*$, where this is not true: above the critical potential, 
the maximum value of the horizontal line can be at b'
in the figure. Therefore, $\vphi$ increases with decreasing $r$
until, at a certain location $r=R$, there are three solutions. The
middle one is an unstable while the other two are stable. At this
point, the composition ``jumps'' between the two stable values and a
discontinuity appears. At such voltages, the profiles are
discontinuous and the coexistence between two distinct phases
occurs.

Assuming that
the jump in $\phi$ occurs at the binodal values, one obtains the
stability criterion \cite{TTL_nature2004}
\begin{equation}\label{DT_wedge}
\Delta T=\frac{v_0}{2k_B}\left|\frac{\Delta\eps}{\phi_c-\phi_0}\right|E^2
~~.
\end{equation}
Here $E=V/(\beta R_1)$ is the largest value of the field.
A mixture of initial homogeneous composition
$\phi_0$ is unstable and demixes into two coexisting domains under
the given field if the temperature is below $T_t+\Delta T$, where
$T_t(\phi_0)$ is the zero-field transition (binodal) temperature at
composition $\phi_0$. In contrast to uniform fields, where field
variations result from composition variations, here field gradient
are due to the non-flat geometry of electrodes. Hence, $\Delta T$
above is typically $2$-$100$ times larger than $\Delta T$ in uniform
fields [Eq. (\ref{DT_landau_mechanism})]. Note that similar demixing
is also expected to occur in a rapidly rotating centrifuge. In that
case $(\omega r)^2$ is the analogue of the spatially-dependent field
$E^2$, where $r$ is the distance from the rotation axis and $\omega$
the angular frequency. The density difference
$\Delta\rho\equiv\rho_2-\rho_1$ replaces the dielectric contrast
$\Delta\eps$ \cite{tsori_crphysique2007}.

Eq. (\ref{DT_wedge}) may be inverted to give the critical voltage for demixing
$V^*$ as a function of $\phi_0$ and temperature. One finds
that $V^*\propto (T-T_t)^{1/2}$. In the experiments of the Leibler group,
conducted using sharp ``razor-blade'' electrodes, the measured exponent was
$0.7\pm 0.15$, larger than the value $1/2$ cited here. 
One may write the dimensionless potential as $U_w\equiv V
\left[v_0\eps_0/(4\beta^2k_BT_cR_1^2)\right]^{1/2}$, 
where $\eps_0$ is the vacuum permittivity.
The critical value of $U_w$ for a closed wedge, $U_w^*$, is
obtained by an approximation similar to that of Eq.
(\ref{DT_wedge}), namely \cite{marcus_jcp2008}
\begin{eqnarray}
U_w^{*2}=\frac{v_0}{k_BT_c}\frac{\phi_t-\phi_0}{4|\Delta\eps|/\eps_0}
\frac{d^2f_m(\phi_t)}{d\phi^2}g(x)~~,
\end{eqnarray}
where $\phi_t$ is the transition composition, $x=R_2/R_1$, and the
dimensionless function $g$ is given by $g(x)=2(x^2-1)/(x^2-1-2\ln x)$.

~\\
{\bf Dynamics}\\
The phase ordering dynamics of mixtures in electric field is quite different
from the no-field case, since the electric field introduces a preferred
direction and thus breaks the initial system symmetry. The phase transition
studied here is even more difficult, because spatially nonuniform fields also
break the translational symmetry.

In this phase transition, droplets nucleate everywhere, not only in regions of
high electric field. As they grow, they move under the external force.
The viscosity
plays an important role, in addition to the field's amplitude, location in the
phase diagram and distance from the binodal, and dielectric constant mismatch
$\Delta\eps$.
Clearly, the spatial dependence of the electric field means the
initial destabilization and phase ordering dynamics are quite different
from the well-studied normal-field instability in thin liquid
films
\cite{hermin1999,russell_steiner2000,russell2003,russel2003,tsori_rmp2009}
and the
regular coarsening dynamics \cite{onuki_book,bray}.

For salt-free mixtures, the starting point for the dynamics
is the following set of equations \cite{onuki_pre2004,tanaka,bray}:
\begin{eqnarray}
\frac{\partial\phi}{\partial t}+{\bf u}\cdot\nabla\phi&=&L\nabla^2
\frac{\delta f}{\delta\phi}~~,\label{dynam_1st} \\
\nabla\cdot(\eps(\phi)\nabla\psi)&=&0~~,\label{dynam_2nd}\\
\nabla \cdot{\bf u}&=&0~~,\label{dynam_3rd}\\
\rho\left[\frac{\partial {\bf u}}{\partial t}+({\bf u}\cdot{\bf \nabla}){\bf
u}\right]&=&\eta\nabla^2{\bf u}-\nabla P-\phi\nabla\frac{\delta f}{\delta\phi}~~.
\label{dynam_4th}
\end{eqnarray}
${\bf u}$ is the velocity field corresponding
to hydrodynamic flow and $\eta$ is the liquid viscosity. Equation
(\ref{dynam_1st}) is
a continuity
equation for $\phi$, where $-L\nabla(\delta f/\delta \phi)$ is the diffusive
current due to inhomogeneities of the chemical potential, and $L$ is the
transport coefficient (assumed constant). Equation (\ref{dynam_2nd}) is
Laplace's
equation, Eq. (\ref{dynam_3rd}) implies incompressible flow, and Eq.
(\ref{dynam_4th}) is
Navier-Stokes equation with a force term $-\phi\nabla\delta
f/\delta\phi$ \cite{onuki_book,bray}.
It should be noted that similar models have been proposed in the
literature; the main differences here are the bistability of $f_m$ and the
nonuniform fields derivable from the potential $\psi$.
The presence of salt is naturally incorporated into the model by adding two
continuity equations for the two ionic species, and by using Poisson's equation
instead of Laplace's.
As a starting point, we assume there is no net flow due to pressure
gradients or
moving solid surfaces -- flow will be purely a result of the forces exerted by
the electric field. In addition, the liquid viscosity $\eta$ is taken as a simple
constant scalar, independent of mixture composition.

Since the equations are coupled and nonlinear, it is useful to first
study the demixing in one of the geometries mentioned above (sphere,
cylinder, or wedge). Consider, for example, the simplifications of
the phase-ordering equations occurring in the system of concentric
cylinders. In this annular capacitor, the no-flow conditions on the
inner and outer cylinders lead to a vanishing flow velocity: ${\bf
u}\equiv 0$ everywhere. One is therefore left with only a single
equation to solve, $\partial\phi/\partial t=L\nabla^2 \delta
f/\delta\phi$. This equation can be viewed as a continuity equation
$\partial\phi/\partial t=-\mathbf{\nabla}\cdot\mathbf{J}$, with a 
current density $J=-L\nabla\delta f/\delta\phi$. In a closed system, {\bf
J} vanishes at $R_1$ and $R_2$, and the integral
$\int^{R_2}_{R_1}2\pi r \phi(r,t) {\rm d}r$ is kept constant
throughout the system dynamics. Gauss's law readily gives the
electric field in the concentric capacitor when the charge is given.
In the explicit scheme we used, $\phi(t)$ is given by a successive
summation of $-(\mathbf{\nabla}\cdot\mathbf{J})dt$, calculated for
each time interval $dt$. The initial condition for the calculation
was a homogeneous distribution $\phi_0$.

\begin{figure}[h!]
\begin{center}
\includegraphics[scale=0.5,bb=50 175 540 790,clip]{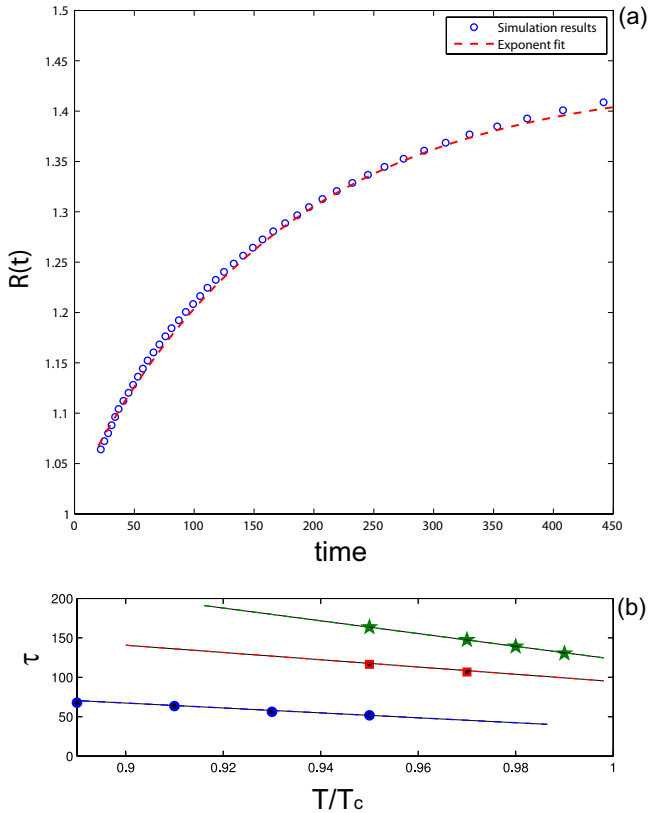}
\end{center}
\caption{(a) Plot of the $R(t)$, the dynamics of front location
between coexisting phases, for a mixture confined by two concentric
cylinders when the charge on the inner cylinder is above the
demixing threshold. $\phi_0=0.3$, $T/T_c=0.95$, and the
dimensionless charge of the inner cylinder is $U_c=0.445$, where
$U_c\equiv \sigma\left[v_0/(4\eps_0k_BT_c)\right]^{1/2}$,
$\sigma$ is the charge per unit area of the cylinder, and $R_1$
is its radius. The exponential time constant $\tau$ in Eq.
(\ref{exp_time_depend}) is plotted in (b) for different values of
$\phi_0$ and $U_c$. Blue circles: $\phi_0=0.2$, $U_c=0.252$. Green
stars: $\phi_0=0.3$, $U_c=0.199$. Red squares: $\phi_0=0.3$,
$U_c=0.252$. $R$ is scaled by $R_1$, and time is scaled by
$Lk_BT/(v_0R_1^2)$. } 
\label{fig_dynamics}
\end{figure}

When the charge on the capacitor is
larger than a threshold charge, we observe fast creation of a
discontinuity near the inner cylinder which then starts to move
outsides. The location of this front, separating
the inner and outer regions, $R(t)$, is shown against time in Fig.
\ref{fig_dynamics} (a). In closed systems, $R$ cannot grow
indefinitely, since mass conservation dictates an upper bound
$R_{\rm max}$ given by
\begin{equation}
R_{\rm max}^2=\left(R_2^2-R_1^2\right)\phi_0+R_1^2~~.
\end{equation}

In the numerical calculation, we find a match to an exponential
relaxation with a single time constant $\tau$:
\begin{equation}\label{exp_time_depend}
R(t)=R_{t=0}+(R_\infty-R_{t=0})(1-e^{-t/\tau})~~.
\end{equation}
$R_\infty$ corresponds to the
steady-state solution; it tends to $R_{\rm max}$ when the voltage or charge
tend to infinity. The time constant $\tau$ depends on the external
potential (or charge) and on the temperature and composition.
Part (b) plots $\tau$ as a function of temperature for different average
compositions. The calculations indicate faster dynamics
(smaller $\tau$) when the average composition is farther from the critical
value (large $|\phi_0-\phi_c|$) or when the cylinder's charge is large, see
Fig. \ref{fig_dynamics} (b).

\section{Mixtures containing salt}

Mixtures of polar liquids (e.g. aqueous solutions)
contain some amount of charge carriers.
In such mixtures, the physics is rich and quite different from the simple
dielectric case. The most important feature is due to screening, occurring
when dissociated ions accumulate at the charged surfaces. This
means that the electric field is substantial only close to the
surfaces, within the screening distance $\lambda$. Field gradients thus
originate from both geometry and screening, and the phase transition
depends on at least two lengths. The ionic screening therefore adds to the
dielectrophoretic
force which separates the liquids components from each other. Since screening is
omni-present, phase-separation may occur even near parallel and flat charged
surfaces, i.e. in one dimension.
But ions have another effect besides increasing the dielectrophoretic force.
Ions have in general different solubilities in the different liquids.
As an ion drifts toward the electrode, it might ``drag'' with it the preferred
liquid component \cite{onuki_jcp2004,onuki_pre2006}. Thus, the solubility
introduces a force
of electrophoretic origin, proportional to the ions' charge.

We use the following free energy density to describe the system:
\begin{eqnarray}\label{FE}
f&=&f_m(\phi)-\frac12\eps(\phi)\left(\nabla\psi\right)^2
+\left(n^+-n^-\right)e\psi\\
&+&k_BT\left[n^+\ln\left(v_0n^+\right)+
n^-\ln\left(v_0n^-\right)\right]-\mu \phi\nn\\
&-&\left(\Delta u^+n^++\Delta u^-n^-\right)\phi-
\lambda^+n^+-\lambda^-n^-\nn
+const.
\end{eqnarray}
The free energy depends on three fields: the electric potential $\psi({\bf r})$,
and the two number densities of positive and negative ions: $n({\bf
r})^\pm$.
The new terms added here are the interaction of ions with the potential
($n^\pm\psi$) and the ideal-gas entropy of ions (logarithmic terms). In
addition,
the parameters $\Delta u^+$ and $\Delta u^-$ measure the affinity
of the positive and negative ions toward the liquid-1 environment, respectively
\cite{TL_pnas2007}. $\Delta u^+$, for example, measures how much a positive
ions prefers liquid-2 environment over that of liquid 1.
$\lambda^\pm$ and $\mu$ are the Lagrange multipliers
(chemical potentials) of the positive and negative ions and liquid
composition, respectively, and $e$ is the electron charge.

The free energy is extremized with respect to the fields $\phi$,
$\psi$, and $n^\pm$:
\begin{eqnarray}
\frac{\delta f}{\delta\phi}&=&\frac{\delta f_m}{\delta\phi}
-\frac12\frac{\delta \eps}{\delta\phi}\left(\nabla\psi\right)^2-\Delta
u^+n^+-\Delta u^-n^--\mu=0\nn\\
\label{eq_gov_eq1}\\
\frac{\delta f}{\delta\psi}
&=&\nabla\left(\eps(\phi)\nabla\psi\right)+e\left(n^+-n^-\right)=0
\label{eq_gov_eq2} \\
\frac{\delta f}{\delta n^\pm}&=&\pm e\psi+k_BT\left(\ln
(v_0n^\pm)+1\right)-\Delta
u^\pm\phi-\lambda^\pm=0\nn\\
\label{eq_gov_eq3}
\end{eqnarray}
in keeping with a fixed mixture and ion concentrations:
\begin{eqnarray}
\mathcal{V}^{-1}\int \phi({\bf r}){\rm d}^3r=\phi_0\\
\mathcal{V}^{-1}\int n^\pm({\bf r}){\rm d}^3r=n_0
\end{eqnarray}
Here $\mathcal{V}$ is the
total volume and $n_0$ the average ion concentration. The Poisson-Boltzmann
equation is
obtained from substitution of Eq. (\ref{eq_gov_eq3}) in the
Poisson equation, Eq. (\ref{eq_gov_eq2}).

Due to these forces, the phase transition is expected to be greatly
enhanced compared to the no-ions case:  it should occur at elevated temperature
above the binodal, and lead to a very thin demixing layer around the charged
object \cite{TL_pnas2007}.
Consider a mixture in the semi-infinite space $x>0$ confined by one wall at $x=0$ charged
at potential $V$.
An approximate formula for the temperature below which a phase transition occurs,
$T_t+\Delta T$, can be obtained by performing a first loop in a perturbative solution
of the equations, namely using a uniform dielectric constant in Eqs. (\ref{eq_gov_eq2})
and (\ref{eq_gov_eq3}) and substitution in Eq. (\ref{eq_gov_eq1}). The expression for
$\Delta T$ is then found to be \cite{TL_pnas2007}:
\begin{eqnarray}\label{DT_ions}
\frac{\Delta T}{T_c}=\left(\frac{|\Delta\eps|}{\eps_c}+\frac{\Delta u}{k_BT_c}\right)
\frac{n_0v_0}{\phi_0-\phi_c}\exp\left(-\frac{e V}{k_BT_c}\right)
\end{eqnarray}
Here $\Delta u=|\Delta u^\pm|$. In most cases, $\Delta\eps/\eps_c\sim \Delta u/k_BT\sim 1$,
and therefore the dielectrophoretic and solubility forces have the same magnitudes. The
numerator $n_0v_0$ is quite small: if we take $v_0=8\times 10^{-27}$
m$^{3}$ and average ion density $n_0=6\times 10^{19}$ m$^{-3}$ ($10^{-7}$M)
we get $n_0v_0\simeq 5\times 10^{-7}$. However, $\Delta T$ is usually large. Even if
we ignore the denominator $|\phi_c-\phi_c|^{-1}$, the exponential factor can be huge: if the
surface potential is only $1$ Volt and $T_c$ is the room temperature, we get
$eV/k_BT_c\simeq 40$. This shows us that demixing should be observed even if the surface
potential $V$ or the charge density $n_0^\pm$ are much smaller.

We would like to note that for homogeneous dielectric liquids, the equation
$\nabla(\eps\nabla\psi)=0$ means that an increase of the potential on the bounding surfaces
simply increases the potential $\psi$ proportionally, 
but that for ion-containing mixtures this
is not true: due to the nonlinearity of the problem, increase of the external potential leads
to a change in the whole distribution $\psi({\bf r})$. The composition difference between
coexisting phases increases with $V$, and the front separating the domains may move to larger
or smaller radii.

\section{Conclusions}

The steady-state and dynamics of phase transitions due to inhomogeneous electric
field are discussed.
In nonpolar mixtures,
the composition profile of a mixture is given for three geometries with azimuthal
or spherical symmetries. Above $T_c$, the profile is smooth, while below $T_c$ it becomes
discontinuous if the surface charge or voltage exceed their critical values.
The location of the front separating the two coexisting phases in equilibrium moves
to larger values of $r$ as the charge or potential increase.
In the restricted case shown here, the main feature of the dynamical process towards
equilibrium is the exponential relaxation of the front location. The exponential time
constant
decreases when the potential is diminished or when the distance from the critical
composition is reduced.

When salt is present, the phase transition is enhanced because
ionic screening leads to a dielectrophoretic force. In addition, the ions' solubility
leads to a strong force of electrophoretic origin. Thus, the transition is g
strengthened and should occur at virtually all temperatures above the binodal 
even at modest
salt content. Our results nicely complement the recent studies by Onuki and co-workers 
on the solvation effects of ions in near critical mixtures 
\cite{onuki_epl1995,onuki_pre2004,onuki_jcp2004,onuki_pre2006}.

A
similar phase transition was observed for a monolayer of surfactant
mixture subject to an electric field emanating from a charged wire
passing perpendicular to the monolayer \cite{KYL}. The more polar
surfactant was attracted to the wire when the field was applied,
while the less polar surfactant was repelled. The effect observed
was linear in electric field because (i) the dipoles were fixed and
not induced, and (ii) they were confined to a plane and could not
twist up-side-down when the field's polarity was reversed.

We point out that when charged colloids are dispersed in aqueous solutions, a thin wetting
layer could be formed due to field-induced demixing, depending on the average salt content,
temperature, and colloid charge. According to Eq. (\ref{DT_ions}), this demixing is quite
favorable, and one needs not be very close to the binodal curve. This should have implications
on colloidal aggregation \cite{beysens} and on the interaction between charged surfaces in
solution \cite{bechinger_nature2008}, because the electrostatically-induced capillary
interaction between the surfaces is expected to be attractive.

\section*{Acknowledgment}

We thank L. Leibler and F. Tournilhac for help in developing
the ideas presented in this work.
This research was supported by the Israel Science
foundation (ISF) grant no. 284/05, and by the German Israeli Foundation (GIF) grant
no. 2144-1636.10/2006.


\begin{thebibliography}{99}
\bibitem{footnote} One may write
$\mathbf{E}_1$ as a sum:
$\mathbf{E}_1=\alpha_\parallel \mathbf{E}_0+\alpha_\perp
\mathbf{E}_{1\perp}$, where
$\mathbf{E}_{1\perp}$ is the field's component in the direction perpendicular
to
$\mathbf{E}_0$. If $\mathbf{E}_0$ is uniform, it can be shown that
$\alpha_\parallel$ is independent of $\mathbf{r}$, and hence
$\int \eps_c \mathbf{E}_0 \cdot \mathbf{E}_1\vphi~ {\rm d}^3r=0$.
\end{thebibliography}
\end{document}